\documentclass{aa}             
\usepackage{amssymb,graphicx}  

\begin{document}

\thesaurus{}

\title{On the source of viscosity in cool binary accretion disks}
\author{F. Meyer, E. Meyer-Hofmeisterr}

\institute{Max-Planck-Institut f\"ur Astrophysik, Karl
Schwarzschildstr.~1, D-85740 Garching, Germany}

\offprints{Emmi Meyer-Hofmeister}

\date{Received:s / Accepted:}

\maketitle

\begin{abstract}
We suggest that the low viscosity in close binary accretion disks 
during quiescence is due to magnetic fields from the companion star.
In very late evolutionary phase the companion stars become brown
dwarfs and have cooled down to such low a temperature that this
process cannot work anymore. The extremely low viscosity in WZ Sge
stars supports this 
connection between companion stars and viscosity. We further suggest
that magnetic activity in such very cool stars is cut off by their
poor electric conductivity.


\end{abstract}

\section{Introduction}
Gammie \& Menou (1998) have suggested that the low magnetic Reynolds
numbers in the cool state of dwarf novae (DN) accretion disks do not allow
any dynamo action and that this might explain the difference between
high $\alpha$ values in the hot (outburst) state and low $\alpha$
value in the cool (quiescent) state, (viscosity
parametrization according to Shakura \& Sunyaev 1973). This result
leaves open the question where the friction in the cool state comes
from. Our investigation complements the work of Gammie \& Menou (1998).
Based on the analysis of the very long recurrence times (decades) and
very high outburst amplitudes of WZ Sge systems as compared to
regular DN systems of similar orbital period we argue that the cause
for the friction in low state cannot lie in the accretion disks or the
primaries of these systems as they are basically the same, but must be
sought in the nature of the secondary stars. We suggest that the
standard low $\alpha$ friction in cool accretion disks is due to
magnetospheric fields from the secondary. Such fields may be
akin to apparent dynamo action of the sun visible
as ``magnetic carpet'' (Schrijver et al. 1998). During the secular
evolution the
companion star loses mass and cools down. We argue that one
can understand the very low value of the $\alpha$ parameter
derived from simulations of outburst cycles for WZ Sge stars (1/10
of the standard cool disk values for DN) as due to
poor conductivity for the very low temperatures of the late type
secondary stars.

This picture supports the long supposed magnetic nature of friction in
cataclyscmic variable (CV) accretion disks. It does support Gammie \& Menou's
explanation of the difference between outburst and quiescent $\alpha$
values. It further shows that quiescent disks without companion stars
would need some other source of magnetic field entrainment or need a
non-molecular cause of friction.

In the following Sects. 2 and 3 we discuss the low-mass companion stars
and their magnetic activity. The
consequences for the accretion disks are discussed in Sect. 4. In
Sect. 5 we discuss the cut off of chromospheric and coronal activity by
low conductivity. The observations for systems in the late
evolutionary state support our suggestion (Sect. 6).

\section{ Evolutionary models for low-mass stars}
The evolution of low-mass stars has been studied
extensively by D'Antona \& Mazzitelli (1982) including
the effect of mass loss (occurring during the secular evolution
of dwarf novae, Warner 1995). These computations already
showed that the stars become very cool late in their evolution.
New evolutionary models of low-mass stars are based on
the most recent interior physics and the latest generation of non-grey
atmosphere models (Baraffe et al. 1998). The electron degeneracy in
the interior causes a drop of
effective temperature for stars of mass below about 0.2\,$M_\odot$. The 
coolest temperatures of main sequence stars (0.075$M_\odot$) are only
around 2000\,K.
Though mass loss is not included the effective temperature is hardly
affected as confirmed by recent work of Kolb \& Baraffe
(1998). Figure 1 shows the effective temperature reached at the 
final state of evolution as a function of the stellar mass (Baraffe et
al. 1998 and Allard et al. 1996).

Companion stars of very low mass belong to the brown dwarf
regime. They have probably spent a long time ( Gyrs) to arrive at
their present CV period. These might be transitional objects of $M\ge
 0.06M_\odot$ or lithium brown dwarfs $M\le 0.06M_\odot$
or evolved brown dwarfs that have now exhausted
their nuclear fuel and cooled below the coolest main sequence stars
(Allard et al. 1997). 

\begin{figure}[ht]
\includegraphics[width=7.5cm]{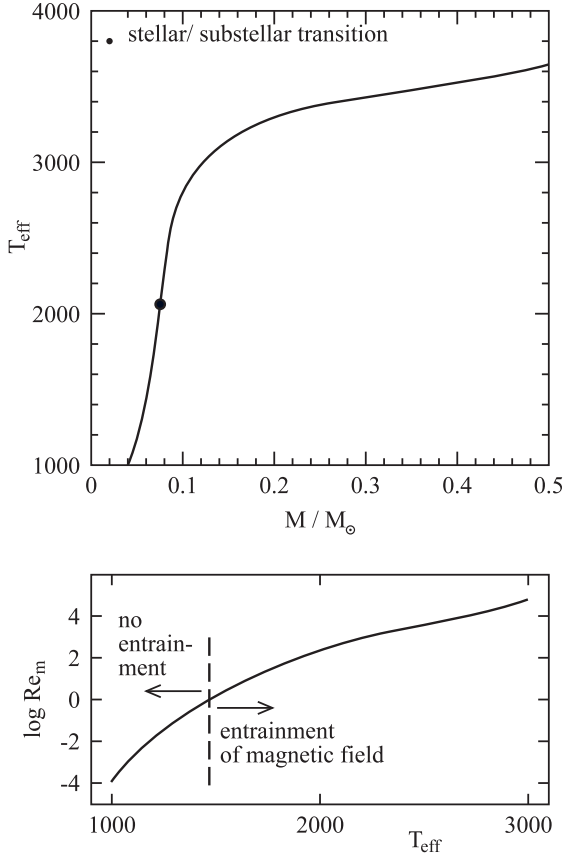}
\caption{Upper panel: Effective temperature of low-mass stars
reached during evolution, from Baraffe et al. (1998) and BD models,
taken at an age of 2 Gyrs
from Allard et al. (1996). Lower panel: Magnetic Reynolds
number of the accretion stream at L$_1$ from a low-mass secondary star.} 
\end{figure}
  
\section{ Magnetic activity of low-mass stars}

The solar cycle dynamo is located at the interface between radiative
interior and convective envelope. Similar cycles on
low-mass companion stars in CVs have been thought to be responsible
for orbital period variations of such systems (Applegate 1992).The
magnetic fields generated by such a dynamo and pulled out with a wind
from the secondary star are thought to be responsible for
the braking of the orbital motion of systems above the period gap.
The disappearance of this dynamo when stars become fully
convective, is commonly accepted as the cause for the occurrence
of the period gap (Spruit \& Ritter 1983, Rappaport et al. 1983).
But also stars of smaller mass, $M \le$ 0.3$M_\odot$, show
chromospheric and coronal activity, indicating magnetic
activity. In the quiet sun such activity is seen in the network of
supergranular cells covering the surface like a ``magnetic carpet''
(Schrijver et al. 1998).

The steady and impulsive heating of chromosphere and corona is
attributed to release of magnetic stresses brought into the field by
twisting and shearing subsurface motion. This activity thus requires
deviations from current-free potential field configurations. For very
low effective temperatures the gas becomes neutral and the electrical
conductivity of the few remaining metal electrons is exponentially
reduced. Thus for very low temperatures one expects that a near-surface
dynamo will stop working and no magnetic fields are
present. Alternatively
a dynamo operating in the bulk of the convection zone will continue to
work beneath the surface. However it is unable to transport magnetic
energy through the non-conductive current-free surface layers. In both
cases observable activity ceases (see evaluation Sect. 5). This seems
to be supported by observations
(for a review of stellar activity in low-mass
stars and brown dwarfs see Allard et al. 1997).

\section{ Magnetic flux entrained in the accretion disk}
The situation described in the preceding section has consequences for
the accretion stream. As long as the secondary star's magnetic field
is present and conductivity is sufficient the magnetic flux pervading
the gas will be brought over to the disk. As soon as the field
disappears or the conductivity has become so poor that the field can
diffuse out of the gas this process ends and the accretion disk is fed
by non-magnetic gas. Thus one expects a significant change of the
magnetic properties of the accreting gas when the secondary stars
become too cool for magnetic activity.

In order to determine whether the gas passing through the
 L${_1}$-point entrains the magnetic field we
have determined a magnetic Reynolds number Re$_m=H \cdot V_s/\eta$,
where $H$ is the length scale, $V_s$ the isothermal sound speed, and
$\eta$ the magnetic diffusivity, all values calculated at the
L${_1}$-point. We take the length as the square root of the cross section $Q$

\begin{equation}
Q = \frac{2\frac{\cal R}{\mu}T a^3}{GM}\frac{1}{6.056}
\end{equation}
where ${\cal R}$ is the gas constant, $\mu$ the molecular weight,$T$ the
temperature, $a$ the distance between the stellar centers, $G$ the
gravitational constant, and M the total mass of the binary. The
numerical factor results for the assumed mass ratio $q = M_2/M_1$ =
1/10, characteristic for 0.7$M_\odot$ primaries and  Roche lobe filling
late main sequence brown dwarf secondaries. For the orbital period
$P_{orb} = 2\pi (a^3/GM)^{1/2}$ we take 80 min. The density $\rho$  in the
flow is obtained from $\dot{M} = Q V_s \rho$,
where we assume a value for the mass transfer rate $\dot{M}$ of
$10^{-11}$$M_\odot/yr$, characteristic for such systems (Kolb 1993). In $V_s
= (\frac {\cal R}{\mu}T)^{1/2}$ we use $\mu = 2.4$
for a mixture of molecular hydrogen and $10\%$ helium. Values of $\rho$
are then of order $10^{-6}$ g/cm$^3$, dependent on the temperature. The
value of the magnetic diffusivity $\eta$ is determined from the
electron density and electron-molecule collision frequency (for cross
sections see Ramanan and Freeman, 1991). See also Gammie and Menou
(1998) for the same procedure for an accretion disk situation. This yields

\begin{equation}
\eta = 10^{3 \cdot 99}T_3^{1/2} n_n/n_e
\end{equation}
where $T_3$ is temperature in units of 1000K and $n_n$ and $n_e$ are number
densities of neutrals and electrons, respectively.
The ratio of number densities can be derived from ionization
equilibrium (Allen 1973) of the electron providing alkali metals. For low
temperatures the contribution of K dominates. We obtain

\begin{equation}
\rm{log} \frac{n_e}{n_n} = 6.48 - \frac{10.94}{T_3} +
\frac{3}{4} \rm{log} T_3 -
\frac{1}{2} \rm{log}\, n_n.
\end{equation}

The number density of neutrals is obtained from the density by division
with the mean particle mass. Putting everything together we obtain

\begin{equation}
\rm{log} Re_m = 7.13 + 2\,\rm{log} T_3 - \frac{10.94}{T_3}
\end{equation}

This function is plotted in Fig. 1. It mirrors the strong temperature
dependence of electron number density at temperatures below the
ionization temperature. At temperatures below 1470K \,the
Reynolds number becomes smaller than 1, indicating that the accretion
stream can no longer entrain any magnetic field of the secondary.

It is remarkable that this temperature divides the very late main
sequence stars from old brown dwarfs that have cooled for $10^9$ to
$10^{9.5}$ yrs. 
Our interpretation suggests that the typical secondaries of WZ Sge type
stars with their long outburst intervals and small $\alpha$-values are
old brown dwarfs consistent with the upper bound of $T_{\rm {eff}}$
from
Ciardi et al. (1998). 

\section{ Suppression of chromospheric and coronal activity for low
temperatures of secondaries}

At the same low temperatures the photospheres of the secondary stars
become extremely poor electrical conductors. This must quench any
magnetic dynamo that operates in surface near layers. It also
drains energy from a dynamo that operating in the depth of the
convection zone and might bring its dynamo number below the critical
value. The magnetic field then would disappear.

We illustrate this poor conductivity by calculating a critical length
scale $l_{\rm{crit}}$ at the surface of the secondaries. 
For motions on
scales smaller than $l_{\rm{crit}}$ the magnetic flux is no longer
carried along by the convective motions. We use data
for photospheric temperature and density kindly
provided by I. Baraffe and F. Allard
Assuming a convective velocity
$v_c$=1 km/s we obtain the following values.

(a) For a low-mass star, $M=0.08M_\odot$,
T$_{\rm{eff}}$=2300K, log$\,\rho$=-4.5:
$l_{\rm{crit}}$= 50\,km. (b) For a brown
dwarf, cooled
to T$_{\rm{eff}}$=1000K, log$\rho$=-4: $l_{\rm{crit}}$= 1.6\,10$^8$ km,
much larger than
the stellar radius. (c) For a young brown dwarf with
T$_{\rm{eff}}$=2700K,
as observed in   X-rays by Neuh\"auser \& Comer\'on (1998) and assumed
log$\,\rho$=-4.7:\,\, $l_{\rm{crit}}$=6\ km. 
The values depend sensitively on the temperature.

Thus stars and brown dwarfs with T$_{\rm{eff}}$ above
about 1600K would be expected to show activity as long as they rotate
sufficiently fast (c.f. Delfosse et al. 1998), while cooler brown
dwarfs might still have magnetic fields from a subsurface dynamo but
could not have magnetically produced activity in their atmospheres
and coronae.

\section{ Indications from observation}
WZ Sagittae is the prototype of DN systems in the evolutionary latest
phase (Osaki 1996). They show rare and luminous outbursts which can only be
understood assuming very low viscosity 
( for recent modeling see Meyer-Hofmeister et al. 1998). 
In Fig. 2 we show the recurrence time of superoutbursts as a function
of the  orbital period. The
recurrence time can be used as an indicator of the viscosity.Thus the
very low $\alpha$ values are expected for the
shortest periods. Note, that systems with regular and with very long
recurrence time both populate this region. This suggests
different temperatures of the corresponding secondary stars. We note that for
extremely low masses in latest evolution the radius of the brown
dwarfs increase with the decreasing mass (Hubbard 1994) and therefore
brown dwarfs and main sequence stars can have the same mean density (that is 
similar periods), but different temperature and thus different
friction in the disks. We point out that near-infrared broad
band photometry has confirmed the low temperature of the secondary
star of WZ Sge, $T_{\rm{eff}}$ less than 1700K (Ciardi et al. 1998).

\begin{figure}[ht]
\includegraphics[width=8.cm]{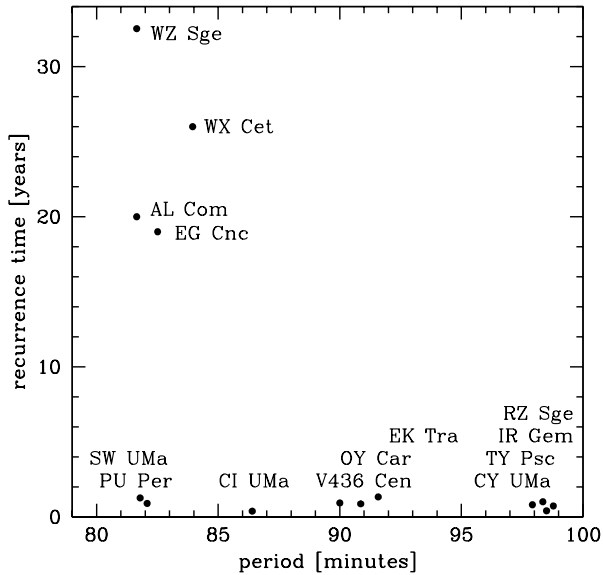}
\caption{Recurrence time of superoutburst of SU Uma stars taken from
Ritter \& Kolb (1998), in addition times for WX Cet (O' Donoghue D. et 
al. 1991) and EG Cnc (Matsumoto et al. 1998)}
\end{figure}

\section{ Conclusions}
\subsection{ Hierarchy of viscosity values}
Why the viscosity during quiescence is so small in WZ Sge stars has
been an open
question addressed by several authors (e.g. Smak 1993, Howell et
al. 1995, Lasota et al. 1995, Osaki
1995, Warner et al. 1996, Hameury et al. 1997). In their work on TOADs
Howell, Szkody and Cannizzo (1995) already raised the question whether
the magnetic flux entrained in the mass stream could stop.
We suggest the following picture: the companion stars have magnetic
activity and their magnetic field passing through the disk or
entrained with the mass flow causes a
viscosity of a few hundredths (in the $\alpha$ parametrization) as
needed for the modeling of quiescent disks, but only if the secondaries
are not too cool. 
The hierarchy of viscosity values can be understood as follows 

\begin{itemize}
\item {In hot accretion disks the small scale dynamo in the
disk can produce a viscosity  of a few tenths ($\alpha$
parametrization). The companion's star magnetic field is not important.}

\item { In cool disks the fields from the secondaries' magnetosphere
penetrating the disk or entrained with the mass stream are the origin
of the viscosity.}

\item { If the secondary is very cool (WZ Sge stars), its magnetic
activity ceases, no fields exist or can be entrained anymore and no
magnetic viscosity is generated in the disk.}
\end{itemize}
 The origin  of the extremely low viscosity is still
unknown, it might be generated by any kind of waves (Narayan et
al. 1987, Spruit et al. 1987, Sawada et al. 1987, Spruit 1987). 
The extremely low viscosity in WZ Sge stars might then be of the same
nature as the viscosity in disks around young stars (FU Orionis
stars) and in AGN disks.
The fact that the outburst of WZ Sge can be modeled with
$\alpha$=0.3 for the hot disk (Osaki 1995) supports the concept that the
viscosity in hot disks is caused in the same way in all dwarf novae.

\subsection{ Cut off of magnetic activity}
The observation of chromospheric and coronal activity in low-mass
stars suggest a dynamo working in the convective region of the
stars. We argue, that magnetic activity in very cool stars is cut off
due to their extremely poor electrical conductivity.

\begin{acknowledgements}
We thank Hans Ritter, Henk Spruit, Nigel O. Weiss and in particular
Isabelle Baraffe for interesting discussions, and Isabelle Baraffe for
providing us with stellar model data.
\end{acknowledgements}

\end{document}